\documentclass[11pt]{article}
\usepackage{amsmath}

\makeatletter
\@addtoreset{equation}{section}
\makeatother

\def\half {\frac{1}{2}}
\def\ben{\begin{equation}}
\def\een{\end{equation}}
\def\bea{\begin{eqnarray}}
\def\eea{\end{eqnarray}}
\def\be{\begin{equation}}
\def\ee{\end{equation}}
\def\br{{\bf r}} 
\def\bx{{\bf x}}
\def\bp{{\bf p}}
\def \bB {{\bf B}}
\def\nn{\nonumber}

\input amssym.def
\input amssym.tex
\def \babla{\boldsymbol \nabla}
\def \bpi{\boldsymbol \pi}
\def\ft#1#2{{\textstyle{\frac{\scriptstyle #1}{\scriptstyle #2} } }}
\def\fft#1#2{{\frac{#1}{#2}}}
\def\del{{\partial}}

\def\rK{{\rm K}}
\def\rH{{\rm H}}
\def\rJ{{\rm J}}

\def\cH{{\cal H}}

\def\turk{{\.{I}n\"on\"u\ }}

\def\bE{{\bf E}}
\def\bC{{\bf C}}
\def\ba{{\bf a}}
\def\bb{{\bf b}}
\def\bu{{\bf u}}
\def \bvom{\boldsymbol \varpi}
\def\dalemb#1#2{{\vbox{\hrule height .#2pt
        \hbox{\vrule width.#2pt height#1pt \kern#1pt
                \vrule width.#2pt}
        \hrule height.#2pt}}}
\def\square{\mathord{\dalemb{6.8}{7}\hbox{\hskip1pt}}}

\def\bequ{\begin{equation}}
\def\eequ{\end{equation}}
\def\barr{\begin{array}}
\def\earr{\end{array}}
\def\bm{\bibitem}

\newcommand{\hoch}[1]{$\, ^{#1}$}

\newcommand{\auth}{\Large\bf{G.W. Gibbons\hoch{\dagger}
and C.N. Pope\hoch{\ddagger,\dagger}}}

\begin{document}

\begin{flushright}
\hfill {
DAMTP-2010-67\ \ \
MIFPA-10-42\ \ \
}\\
\end{flushright}

\begin{center}

{\Large{\bf Kohn's Theorem, Larmor's Equivalence Principle and
the Newton-Hooke Group}}

\vspace{30pt}
\auth

\large

\vspace{10pt}{\hoch{\dagger}\it DAMTP, Centre for Mathematical Sciences,\\
 Cambridge University, Wilberforce Road, Cambridge CB3 OWA, UK}

\vspace{10pt}{\hoch{\ddagger}\it George P. \& Cynthia W. Mitchell
Institute for\\ 
Fundamental Physics and Astronomy,\\ Texas A\& M University,
College Station, TX 77843-4242, USA}

\end{center}

\vspace{30pt}

\begin{abstract}

We consider non-relativistic  electrons, each of the same charge to mass 
ratio,
moving in an external magnetic field  with an  
interaction potential depending only on the mutual separations, 
possibly confined
by a harmonic trapping potential. We show that the system
admits a ``relativity group'' which is a one-parameter family of
deformations of 
the standard Galilei group to the Newton-Hooke group
which is a Wigner-\turk contraction of the de Sitter group.
This allows a group-theoretic  interpretation 
of Kohn's theorem and related results. Larmor's Theorem
is used  to show that the one-parameter family of deformations
are all isomorphic. We study the ``Eisenhart'' or ``lightlike''  lift
of the system, exhibiting it as a pp-wave. 
In the planar case, the Eisenhart lift is the 
Brdi\v{c}ka-Eardley-Nappi-Witten  pp-wave  solution   
of Einstein-Maxwell theory, which may also be regarded as
a bi-invariant metric on the Cangemi-Jackiw group.   
\end{abstract}
\pagebreak
\setcounter{page}{1}
\newpage

\section{Introduction}

Since at least the time of Galieo Galilei, 
physicists have realised the importance of understanding how
physical quantities  change under changes of the
frame of reference. With the development of Special
Relativity, attention became focussed on the underlying
group, and its invariants. Einstein's construction of 
the theory of General Relativity was based on the related
idea that the paths of ``freely falling particles'' should be
universal, that is to say, independent of their mass or other 
properties.   However, as we aim to show 
in this paper,  the utility of this viewpoint
is  not restricted to high energy physics or general relativity, 
but may also
be applied with advantage to situations where the energies and speeds of the
particles are moderate compared with that of light,
and the gravitational fields are weak.
All that matters is that the dynamical behaviour of the
particles should be universal and that some sort of ``Equivalence Principle''
hold.

 These requirements are met in 
an extremely useful and well studied model in condensed matter physics,
consisting  of $N$ electrons, each of mass $m$ and charge $e$, 
moving in a uniform  background 
magnetic field ${\bf B}$. The electrons are assumed to interact with
each other via a potential $V$ that depends only on their $\half N(N-1)$
relative  positions: $V= \sum_{a\ne b} V({\bf r}_a-{\bf r}_b)$, $a,b=1,2, ,N$.
A typical example would be a Coulomb potential, which is actually  central,
i.e. which depends only on the relative distances 
$r_{ab} =|{\bf r}_a-{\bf r}_b |$. However, for what follows $V$ need not
be central. 
In addition, especially in the case of  ``quantum dots'' 
\cite{Yngvason}, the electrons
may be confined by an additional 
harmonic
or ``parabolic''  trapping potential of the form 
$\half m \omega^2 \sum _a {\bf r}_a^2$, or more generally,
an anisotropic oscillator potential. The classical Lagrangian
in the isotropic case 
is therefore
%%%%%
\ben
L= \sum_a \Bigl(  \half m {\dot{\bf r}}_a^2  +e {\bf A}_a \cdot {\dot {\bf
    r}}_a
- \half m \omega^2 \,{\bf r}_a^2 \quad \Bigr)  - 
   \sum_{a\ne b}V({\bf r}_a-{\bf r}_b)\,,   
\label{Lag}
\een
%%%%% 
and the classical equations of motion are
%%%%%
\ben
m \ddot {\bf r}_a = e\dot {\bf r}_a  \times {\bf B}   - m \omega ^2
{\bf r} _a -\babla _a V \,, \label{eoms} 
\een 
%%%%%
where $ \babla_a = \frac{\partial }{\partial {\bf r}_a}$,
 ${\bf A}_a = {\bf A}({\bf r}_a)$ and  ${\bf B} = \babla \times {\bf A}$ .
Since ${\bf B}$ is assumed uniform, ${\bf A}$ may be taken to be
linear in ${\bf r}$, and in what follows we shall adopt the gauge
for which 
%%%%%
\be
{\bf A}({\bf r})= -\ft12 {\bf r} \times {\bf B}\,.\label{gaugechoice}
\ee
%%%%%  

   The system of equations (\ref{eoms}) has an extremely  important property,
first pointed out by Kohn \cite{Kohn} in the case without trapping potential
(i.e. with $\omega=0$ ),  and referred to as {\it Kohn's
Theorem}.  This theorem states that one may split off the ``centre of mass 
motion'' in such a way that the interaction potential 
$V({\bf r}_a-{\bf r}_b)$ does not enter. In other words, the centre of
mass degree of freedom, ${\bf r}= \frac{1}{N} \sum _a {\bf r}_a $,
completely decouples from the $(N-1)$ independent
relative degrees of freedom in a universal way. For this property to
continue to hold when a trapping potential is present,
the potential must be the sum of identical quadratic terms for each electron, 
although each quadratic
term need not be, as it is  in (\ref{Lag}), isotropic in  ${\bf r}$. 

  One way to obtain Kohn's result is to substitute  the identity
%%%%%
\ben
\sum_a   {\dot {\bf r}} _a ^2 = 
\frac{1}{N} \bigl( \sum _a  {\dot {\bf r } } _a  \bigr) ^2
+ \frac{1}{N}   \sum_ {a<b} ({\dot {\bf r}}_a-{\dot  {\bf r}} _b) ^2   
\een
%%%%%
into the Lagrangian (\ref{Lag}).  Alternatively, one may 
note that the  equations of motion (\ref{eoms}) 
 admit a  symmetry
%%%%%
\ben
{\bf r}_a \rightarrow  {\bf r}_a + {\bf a}(t)  
\een
where 
%%%%%
\ben
\ddot {\bf a} = \frac{e}{m}  \dot {\bf a} \times {\bf B} - \omega^2
      {\bf a} 
\label{gal} 
\een
%%%%
For this reason, Kohn's theorem is often ascribed to the 
Galilei invariance of the system. However this is not actually correct
since
the solutions of (\ref{gal}) are not of the form ${\bf a}= {\bf b} +
{\bf u} t $. Moreover, the presence  of a non-vanishing 
 magnetic  field ${\bf B}$  breaks the $SO(3)$ rotational
symmetry down  to the $SO(2)$ subgroup of rotations about the direction
of the magnetic field. Nevertheless, for fixed $\omega$ 
 and $B=|{\bf B}|$,  the second-order differential
equation 
(\ref{gal}) does have a 6-parameter set of 
solutions  
which in general defines a 6-parameter 
symmetry group. This six-dimensional  group of generalised 
translations and boosts
is abelian, just as in the Galilei case. However since the 
transformations, i.e. the solutions of
(\ref{gal}),  depend non-trivially
on time, when time translations are taken into account 
we get a seven-dimensional non-abelian group. 
Thus the relevant symmetry group 
is not the translation,  boost and time-translation
subgroup of the  Galilei group, but appears to  belong instead to 
 a continuous one-parameter 
family of deformations depending on the dimensionless
ratio $eB/(m \omega)$. If the residual rotations are 
added we get an eight-dimensional group. However the apparent dependence
on the dimensionless parameter  $eB/(m \omega)$ is illusory, since
by an application of a well known result of Larmor \cite{Larmor}, 
the magnetic field
$B$ may be eliminated by passing to a rotating frame with angular
velocity $\bvom$ given by
%%%%%
\ben
\varpi= -\frac{eB}{2m} \,,
\een
%%%%%  
at the expense of introducing a quadratic potential. 

To see this in detail, note that if we pass to a frame that is rotating with
constant angular velocity $\bvom$ then for any vector $\bx$,  
%%%%%
\ben
\dot \bx = \bx ^\prime + \bvom \times \bx \,,
\een
%%%%%
where a $\dot{ \,}$  indicates a time derivative of the 
 components of $\bx$  in an inertial or space-fixed frame, 
and a ${\,}^\prime $ denotes a time derivative of the components 
in the rotating frame. 

  The equations of motion become
%%%%%
\ben
m {\bf r}^{\prime \prime} _a = e {\bf r}_a ^{\prime}  \times
\Bigl ( {\bf B} + \frac{2m}{e}  \bvom \Bigl)     - 
m \omega ^2 {\bf r} _a  +   e (\bvom  \times \br_a ) \times \bB 
-    m \bvom \times (\bvom \times \br_ a)    - 
         \babla _a V (\br_a - \br_b  )\,. 
\een 
%%%%%
The equations of motion (\ref{eoms})
 are unchanged in form if we make the replacement
%%%%%
\ben
\bB \rightarrow \bB +\frac{2e}{m} \bvom
\een
%%%%%
and modify the trapping force, a result usually called Larmor's
theorem. 
However 
%%%%%
\ben
(\bvom \times \br _a ) \times \bB = \half \babla _a   (\br_a \cdot \br_a)
(\bvom \cdot  \bB ) - (\br_a  \cdot \bB ) \bvom  
\een
%%%%%
and 
%%%%%
\ben
\babla _a\ \times \bigl ((\br_a \cdot \bB ) \, \bvom  \bigr ) = \bB
\times \bvom\,,
\een
%%%%%
and so the motion in the modified trapping force is 
not conservative unless either $\bvom \times \bB =0$ or   
the motion is perpendicular to the magnetic field, $\br_a
\cdot \bB=0$. In the latter  case, the passage to the rotating frame
is a symmetry relating equivalent systems connected by
%%%%%
\ben
B \rightarrow \tilde B= 
 B +\frac{2m}{e} \varpi  \,\qquad  \omega ^2 \rightarrow \tilde
 \omega^2  =
\omega ^2 -\varpi ^2 - \frac{e}{m}  B  \varpi = \omega^2 -\bigl(
\varpi + \frac{eB}{2m}\bigr) ^2  +\frac{e^2 B^2 }{4 m^2} \,. 
\een  
%%%%%
Note that under Larmor's transformation
%%%%%
\ben
\Omega ^2 = \omega ^2 + \frac{e^2 B^2 }{4 m^2} = \tilde \Omega^2=
\tilde \omega ^2 +  \frac{e^2 \tilde B^2 }{4 m^2} \,,
\een
%%%%%
and so $\Omega ^2$ is unchanged.

   In the planar case, this is just
a modification of the existing quadratic trapping potential,
and therefore one  may always eliminate the magnetic field.  
One may also check that if one wishes, rather than eliminating 
the magnetic field, one may use a Larmor transformation to eliminate
the trapping potential.    

Taking the first option,  the symmetry group of the system is 
now easy to identify.
It is the  {\it Newton-Hooke group} in $(2+1)$ dimensions. 
In $(3+1)$ dimensions the Newton-Hooke group is one of the
possible {\it kinematic groups } first classified by Levy-Leblond and Bacry    
\cite{BacryLeblond,BacryNuits}, which may be regarded as a small-velocity
limit, or Wigner-\turk  contraction \cite{WignerInonu}, of the 
of the anti-de Sitter group $SO(3,2)$. In effect, the trapping potential   
behaves like a universal cosmic attraction associated  with 
a negative cosmological constant $\Lambda = - 3 \omega ^2/c^2$
\cite{GibbonsPatricot}. The Newton-Hooke group
has been applied to  non-relativistic cosmology, 
both classically and at the quantum level\cite{Derome}.
Thus one is tempted to wonder whether
the deformations where the magnetic field is included
may also have a cosmological application to a Goedel-type
rotating universe with
a cosmological term.  

In the planar case the relevant Newton Hooke group is six-dimensional,
and it is a Wigner-\turk contraction of $SO(2,2)$.

One  purpose of the present paper is to explore the
properties of the 
deformed group in more detail, and to use it to
obtain the quantum mechanical spectrum of the centre of mass motion   
for general ${\bf B}$ and $\omega$ directly, by using group theory. 
This was partially done
by Kohn in his original paper \cite{Kohn} in the case 
$\omega=0$, by using an operator method.  The  full
spectrum of a particle in a magnetic field with harmonic potential  
 was obtained by Fock \cite{Fock}  and by Darwin \cite{Darwin} long ago, 
by solving explicitly for the wave functions
and using the properties of the associated Laguerre polynomials.
We shall obtain the spectrum without solving for the radial functions,
by using oscillator methods.

It is well known that at the quantum mechanical level,
 the classical 10-dimensional  Galilei
group enlarges to Bargmann's  11-dimensional 
central extension \cite{Bargmann}.  
 e  find that a similar phenomenon occurs
in our case.  (For an alternative discussion of central extensions of the
$(2+1)$-dimensional Newton-Hooke group, from the point of view of
nonlinear realisations, see \cite{gomisexotic}.) 
  In order to obtain the quantum mechanical symmetry
group
we first turn to a classical Hamiltonian treatment,
obtaining the canonical generators or ``moment maps'' 
of the seven-dimensional symmetry group. It is well known
that the Lie algebra of a transformation group acting
on phase space will not in general coincide with
the Poisson algebra of its moment maps, because of the
possibility of central terms. That is precisely what we find
in our case.  Upon quantisation of the moment maps by replacing
the canonical momentum $\bpi$ by $-i \hbar\babla$
acting on the quantum mechanical wave function, we
indeed find   that it is the   centrally-extended 
 Poisson algebra that applies at the quantum level.           

In the planar case, we obtain in this way a seven-dimensional 
extended group which is isomorphic to  
$\bigl({\cal C}{\cal J }\otimes {\cal C}{\cal J }\bigr ) /N$,  
where ${\cal C}{\cal J}$ is Cangemi and Jackiw's 
central extension of the   
two-dimensional Euclidean group \cite{Cangemi}, $N$ being
the central extension.

  A nice geometric way of understanding the central extensions 
of non-relativistic symmetry groups acting on a $(d +1)$-dimensional 
non-relativistic spacetime is via a ``lightlike
reduction''  of a Lorentzian metric 
on a $(d+2)$-dimensional spacetime  
that admits a covariantly-constant null Killing vector field 
\cite{Duval,GibbonsDuval}.
According to a result of Eisenhart \cite{Eisenhart},
a rather general  class of 
 mechanical systems may be  lifted so that their motion is a ``shadow''
of a null geodesic in $d+2$ dimensions. In our case $d=3N$, and
we carry out the lift and describe some of the properties
of the higher-dimensional  spacetime.
The Eisenhart lift of the centre of mass motion in the planar case
turns out to be a Nappi-Witten pp-wave \cite{Nappi:93},
 which has arisen in string theory 
\cite{duval1,duval2,duval3,Horowitz:95,duval4}.
This particular four-dimensional  pp-wave is homogeneous,  
with the seven-dimensional Newton-Hooke group as its symmetry. 
In fact it may be identified as the group manifold of the
universal cover of the Cangemi-Jackiw group ${\cal C}{\cal J}$.

The organisation  of the paper is as follows.
In section 2 we shall describe the action of the classical relativity
group of our model on its non-relativistic spacetime. 
We construct its infinitesimal generators as vector fields on the
spacetime
and evaluate their Lie brackets and hence determine  
the associated Lie algebra.
In section 3 we provide a brief account of the modifications brought
about by a uniform  electric field. In section 4 we pass to a
classical Hamiltonian treatment and find that the Poisson
algebra of functions on phase space differs  from the classical Lie
algebra  by central terms. In section  5 we pass to the quantum
theory and find that the central terms are  present in the quantum algebra.
We use the algebra to obtain the wave functions and energy
eigenvalues. In section 6 we lift our system from configuration space
to  a Lorentzian spacetime with two extra dimensions admitting a null
Killing vector, for which the classical motion lifts to that
of a null geodesic. In section 7 we identify the light-like lift
in the planar case to be the  Brdi\v{c}ka-Eardley-Nappi-Witten pp-wave
spacetime.

\section{The Action on Non-Relativistic Spacetime}

 In this section we give the solutions of (\ref{gal}) and
use them to obtain the action of the seven-dimensional
group on  the $(3+1)$-dimensional
non-relativistic spacetime ${\cal  M}_{\rm rel}$
whose coordinates are $(t,{\bf r})$.
This allows us to obtain their infinitesimal generators
as vector fields acting on  ${\cal  M}_{\rm rel}$.
The Lie algebra is then obtained by taking Lie brackets.  

  Consider the system (\ref{gal}) with a uniform magnetic field.
%%%%%
Taking $\bB$ to lie in the $z$ direction, and defining
%%%%%
\be
\nu=\frac{eB}{2m}\,,\qquad \Omega=\sqrt{\omega^2+\nu^2}\,,
\ee
%%%%%
where $\nu$ is the Larmor frequency, 
the general solution of (\ref{gal}) is given by
%%%%%
\bea
a_x&=& \alpha_1\, \cos(\Omega+\nu)t + \alpha_2\, \sin(\Omega+\nu)t +
   \beta_1\, \cos(\Omega-\nu)t -\beta_2\, \sin(\Omega-\nu)t\,,\nn\\
a_y&=& \alpha_2\, \cos(\Omega+\nu)t - \alpha_1\, \sin(\Omega+\nu)t +
   \beta_2\, \cos(\Omega-\nu)t +\beta_1\, \sin(\Omega-\nu)t\,,\nn\\
a_z&=& \alpha_3\, \cos\omega t+ \beta_3\, \sin\omega t\,,
\eea
%%%%%
where $\alpha_i$ and $\beta_i$ are 6 arbitrary constants.  Thus we may 
read off the 6 associated generators:
%%%%%
\bea
\rK_1 &=& \cos(\Omega+\nu)t\, \frac{\partial}{\partial x} -
     \sin(\Omega+\nu)t\, \frac{\partial}{\partial y}\,,\nn\\
\rK_2 &=& \sin(\Omega+\nu)t\, \frac{\partial}{\partial x} +
     \cos(\Omega+\nu)t\, \frac{\partial}{\partial y}\,,\nn\\
\rK_3 &=& \cos(\Omega-\nu)t\, \frac{\partial}{\partial x} +
     \sin(\Omega-\nu)t\, \frac{\partial}{\partial y}\,,\nn\\
\rK_4 &=& -\sin(\Omega-\nu)t\, \frac{\partial}{\partial x} +
     \cos(\Omega-\nu)t\, \frac{\partial}{\partial y}\,,\nn\\
\rK_5 &=& \cos\omega t\, \frac{\partial}{\partial z}\,,\nn\\
\rK_6 &=& \frac{\sin\omega t}{\omega}\, \frac{\partial}{\partial z}\,.
\label{Kidef}
\eea
%%%%%
These all commute. If we define also $\rH=\partial/\partial t$, then we 
have the commutators
%%%%%
\bea
{[}\rH,\rK_1{]}&=& -(\Omega+\nu)\, \rK_2\,,\qquad {[}\rH,\rK_2{]}= 
(\Omega+\nu)\, K_1\,,\nn\\
{[}\rH,\rK_3{]} &=& (\Omega-\nu)\, \rK_4\,,\qquad {[}\rH,\rK_4{]}
= -(\Omega-\nu)\, \rK_3\,,\nn\\
{[}\rH,\rK_5{]}&=& -\omega^2\, \rK_6\,,\qquad {[}\rH,\rK_6{]}= \rK_5\,.
\label{HKalg}
\eea
%%%%%

   In the limit when $\omega=0$, the six $\rK_{(\alpha)}$ generators become
%%%%%
\bea
\rK_1 &=& \cos2\nu t\, \frac{\partial}{\partial x} -
    \sin 2\nu t\, \frac{\partial}{\partial y}
\,,\nn\\
\rK_2 &=&\sin2\nu t\, \frac{\partial}{\partial x} +
\cos 2\nu t\, \frac{\partial}{\partial y}
\,,\nn\\
\rK_3 &=& \frac{\partial}{\partial x}\,,
 \qquad K_4 = \frac{\partial}{\partial y}\,,\nn\\
\rK_5 &=& \frac{\partial}{\partial z}\,,\qquad 
\rK_6 = t\, \frac{\partial}{\partial z}\,.
\eea
%%%%%

Note that even if $\omega^2=0$, and we do not 
obtain the two-dimensional Galilei group,

%   One might be concerned that the algebras (\ref{HKalg}) are merely
%relabellings of a single underlying algebra, for example the Galilean algebra.
%To see that this is in fact not the case, and that algebras with different
%values of the ratio $\nu/\omega$ are non-isomorphic, it suffices to 
%exhibit a quantity formed from the structure constants that is invariant
%under the action of $GL(7,{\Bbb R})$ and that depends non-trivially on
%the ratio $\nu/\omega$. The only non-zero structure constants in the algebra
%of the $\rH$ and the $\rK_{(\alpha)}$ are those that can be read off
%from (\ref{HKalg}), of the form $C_H{}^{(\alpha)}{}_{(\beta)}$. This set
%of structure constants is unaltered if any multiple of the $\rK_{(\alpha)}$
%is added to $\rH$, and it changes by a factor if $\rH$ is rescaled. Regarding
%the structure constants as defining a matrix $C$ ($C$ gives the adjoint
%action of $\rH$ on the abelian subalgebra generated by the $\rK_{(\alpha)}$),
%the $GL(7,{\Bbb R})$-invariant quantity
%%%%%
%\be
% \fft{\tr C^4}{(\tr C^2)^2}= \ft12 - \fft{\omega^2}{4\nu^2+3\omega^2}\,,
%\ee
%%%%%
%which is monotonic in $\nu/\omega$ and therefore establishes the 
%inequivalence of the algebras with different values of $\nu/\omega$.

The equations of motion are invariant under rotation
about the direction of the magnetic field ${\bf B}$. 
This is generated by
%%%%%
\ben
\rJ= x \frac{\partial}{\partial y } - y \frac{\partial}{\partial x} \,. 
\een 
%%%%%
Clearly $\rJ$ commutes with $\rH$, $\rK_5$ and $\rK_6$. It acts by 
rotations on $(\rK_1,\rK_2)$ and
$(\rK_3, \rK_4)$: 
%%%%%
\bea
\bigl [\rJ, \rK_1] &=& -\rK_2\,\qquad \bigl [ \rJ, \rK_2 \bigr ] = \rK_1\nn\\
\bigl [\rJ, \rK_3] &=& -\rK_4\,\qquad \bigl [ \rJ, \rK_4 \bigr ] = \rK_3 \,.
\label{JKcom}
\eea
%%%%%

The six-dimensional sub-algebra spanned by $(\rH,\rJ,\rK_1,\rK_2,\rK_3,\rK_4)$
appears to depend on the two constants $\nu$ and $\Omega$, but as
mentioned earlier, by Larmor's theorem,  the apparent dependence on 
$\nu$ is illusory.   To see this explicitly we may pass to coordinates 
$\tilde x,\tilde y$ in a  rotating frame by setting
%%%%%
\ben
x+iy= e^{i\varpi t} (\tilde x +i\tilde y ) \label{coord}
\een
%%%%%
so that 
%%%%%
\ben
a_1+ia_2 = e^{i\varpi t} (\tilde a _1 +i \tilde a_2 ) 
\een
%%%%%
and find that in the new coordinates $\Omega$ is unchanged but $\nu $ is 
replaced by $\nu + \varpi$. By an appropriate choice of
the angular velocity $\varpi$ of our rotating frame 
we may give the Larmor frequency  $\nu$ any value we wish.
Despite this, we  we shall retain both parameters
in the formulae that follow.

\section{Electric Fields}

We shall now briefly review the situation
when a uniform  time-independent electric field $\bE$  is present,
since this introduces new some features. The equation of motion
is now 
%%%%%
\ben
\ddot {\bf r}_a = \frac{e}{m}  {\bf E} +  \frac{e}{m}
 \dot {\bf r}_a  \times {\bf B}   -  \omega ^2 {\bf r} _a
 -\frac{1}{m}   \babla _a V
\label{eleceom}
\een
%%%% 
The point is that while (\ref{eleceom}) is is still invariant
under the deformed Galilei group, i.e. under
%%%%%
\ben
\br_a \rightarrow \br_a + \ba(t) \,,
\een 
%%%%%
one may enhance the symmetry group  of the equation if one allows
a transformation of the electric field $\bf E$.
Consider, to begin with, the case with  an electric field
but no magnetic field and no trapping potential,  so that
%%%%%
\ben
\ddot {\bf r}_a = \frac{e}{m}  {\bf E}-\frac{1}{m}   \babla _a V \,.
\een
%%%%%
Because  the ratios of charge to mass of all the particles
are identical, the equation of motion is now invariant
if  one transforms $\br_a$ as 
%%%%%
\ben
\br_a(t) \rightarrow  \br_a(t) + \half \bC t^2 
\een
%%%%%
and one transforms the electric field as 
%%%%%
\ben
\bE \rightarrow \bE -\frac{m}{e} \bC \,. 
\een
%%%%

   Indeed if one picks $\bC= \frac{e}{m}\bE$, one may eliminate
the electric field altogether. Clearly this is the analogue 
of Einstein's equivalence principle. The additional
vector fields are
\ben
{\rm C}_i = \half t^2  \nabla _i 
\een
and 
\ben
[ H, C_i]=  t\nabla _i\, \label{boost} 
\een
where the right-hand side of (\ref{boost})  is a Galilean boost.

The inclusion of a magnetic field
changes things. The equation of motion is now 
%%%%%
\ben
\ddot {\bf r}_a = \frac{e}{m}  {\bf E} + 
\frac{e}{m} {\dot \br} _a \times \bB -\frac{1}{m}   \babla _a V \,,
\een
%%%%%
and a boost 
%%%%%
\ben
\br_a(t) \rightarrow \br_a(t) + \bu t 
\een
%%%%%
induces a ``non-relativistic Lorentz  transformation'' 
%%%%
\ben
\bB \rightarrow \bB \,,\qquad \bE \rightarrow \bE - \bu \times \bE    
\,.
\een
%%%%
Indeed if we choose  the drift velocity ${\bf v}_d$ to satisfy  
%%%%
\ben
{\bf E} + {\bf v}_d \times {\bf B} =0\,,
\een
%%%%
we can eliminate the electric field altogether.
This is essentially the classical Hall effect.

The presence  of the trapping potential further complicates  matters
since a translation
%%%%%
\ben
\br_a(t) \rightarrow \br_a(t) + \bb 
\een
%%%%%
induces the transformation
%%%%%
\ben
\bE \rightarrow \bE- \frac{m \omega ^2}{e} \bb  
\een
%%%%%
and if we choose  
%%%%%
\ben
\bb = \frac{e}{m \omega^2} \bE \,,
\een
%%%%%
we can also eliminate the electric field altogether.

The upshot of this discussion would seem to be that
by a suitable symmetry transformation we can always eliminate
any uniform  electric field and we have seen above that we may remove
any uniform magnetic field. 
In what follows we shall not consider electric fields further.

\section{Classical Hamiltonian Treatment}

Starting from
the Lagrangian (\ref{Lag}), 
and the canonical momentum
%%%%%
\ben
{\bpi }_a= m {\dot {\bf r}} _a + e {\bf A}_a = {\bf p}_a+ e {\bf A}_a  \,,
\een
the Poisson brackets are
\ben
\{x_{ia}, \pi_{jb} \}= \delta_{ij} \delta_{ab} \,, \qquad 
\{x_{ia},x_{jb} \}=0\,,\qquad \{\pi_{ia},\pi_{jb} \}=0\,,
\een
%%%%%
and 
%%%%%
\ben
\{x_{ia}, p_{jb} \}= \delta_{ij} \delta_{ab} \,, \qquad 
\{ p_{ia},    p_{ib}   \} = e\delta _{ab}
 \epsilon _{ijk} B_k  \,.
\een
%%%%%
The Hamiltonian is given by
%%%%%
\bea
H&=& 
\frac{1}{2m} \sum_a \bigl(  { \bpi}_a-e{\bf A}_a ) ^2
 + \ft12 m\omega^2\, \sum_a {\bf r}_a^2  + \sum _{a<b} V(\br_a-\br_b )\,,\nn\\ 
&=& \frac{1}{2m} \sum_a   {\bf p }_a  ^2
  + \ft12 m\omega^2\, \sum_a {\bf r}_a^2 + \sum _{a<b} V(\br_a-\br_b )\,,\nn\\
&=& \cH + \cH_{\rm rel}\,,
\eea
%%%%%
where the centre-of-mass and relative Hamiltonians are given by
%%%%%
\bea
\cH &=& \fft1{2M}\, \bp^2 + \ft12 M \omega^2\, \br^2\,,\label{comH}\\
\cH_{\rm rel} &=& \sum_{a,b}\Big(\fft1{2mN} (\bp_a-\bp_b)^2 
 +\fft1{2N}\, m\omega^2  (\br_a-\br_b)^2 +
   V(\br_a-\br_b)\Big)\,,
\eea
%%%%%
$N$ is the number of electrons, and $M=m N$ is their total mass.
Note that we have defined the centre-of-mass coordinates $\br$, and the
centre-of-mass momentum $\bp$ by
%%%%%
\be
\br = \fft1{N} \sum_a \br_a\,,\qquad
\bp= \sum_a \bp_a\,.
\ee
%%%%%
Similarly, we define the centre-of-mass canonical momentum by 
%%%%%
\ben
{ \bpi} = \sum_a { \bpi _a} 
\een
%%%%%
and so we shall have the Poisson bracket relations
%%%%
\ben
\{ { x}_i, {\pi}_j \} =\delta _{ij} \,,\qquad \{ x_i, x_j\}=0\,,\qquad
\{\pi_i,\pi_j \}=0\,.  
\een
%%%%
We also have
%%%%%
\ben
\{ { p}_i, {p }_j \} = e N \epsilon_{ijk} B_k \,, 
 \qquad \{x_i , p_j \} = \delta _{ij}  \,, 
\een
%%%%%
and
%%%%%
\ben
{\dot {\bf p}} = \frac{e}{m}   {\bf p}   \times
  {\bf B} \,.  
\een
%%%%%
Note that $eN$ is the total charge.

  In the ``symmetric'' gauge (\ref{gaugechoice}) that we are using, we have
%%%%%
\be
p_x = \pi_x + \nu M y\,,\qquad p_y= \pi_y -\nu M x\,,\qquad
    p_z= \pi_z\,.
\ee
%%%%%
The generators $\rK_{(\alpha)}$ defined in (\ref{Kidef}) act on the 
$x^i$ and $p_i$, according to
%%%%
\be
\delta_{(\alpha)} x^i = \rK_{(\alpha)}^i\,,\qquad
\delta_{(\alpha)} p_i = m \delta\dot x^i = M \dot \rK_{(\alpha)}^i\,,
\ee
%%%%%
where the dot means a derivative with respect to $t$.  We may therefore
read off the corresponding moment maps $\kappa_{(\alpha)}$, i.e. the functions
on phase space such that
%%%%%
\be
\delta_{(\alpha)} \,x^i = \{x^i,\kappa_{(\alpha)}\}\,,\qquad 
\delta_{(\alpha)} \,p_i = \{p_i,\kappa_{(\alpha)}\}\,.
\ee
%%%%%
We find that they are given by
%%%%%
\bea
\kappa_1 &=& (\pi_x + M\Omega y)\cos(\Omega+\nu)t - 
   (\pi_y-M\Omega x) \sin(\Omega+\nu) t\,,\nn\\
\kappa_2 &=& (\pi_x + M\Omega y)\sin(\Omega+\nu)t + 
   (\pi_y-M\Omega x) \cos(\Omega+\nu) t\,,\nn\\
\kappa_3 &=& (\pi_x - M\Omega y)\cos(\Omega-\nu)t + 
   (\pi_y+M\Omega x) \sin(\Omega-\nu) t\,,\nn\\
\kappa_4 &=& -(\pi_x - M\Omega y)\sin(\Omega-\nu)t +
   (\pi_y+M\Omega x) \cos(\Omega-\nu) t\,,\nn\\
\kappa_5 &=& \pi_z \, \cos\omega t + M\omega z\, \sin\omega t\,,\nn\\
\kappa_6 &=& \pi_z\, \fft{\sin\omega t}{\omega} - M z\, \cos\omega t\,.
\eea
%%%%%

   It is straightforward to see that the moment maps have the 
following non-vanishing Poisson brackets:
%%%%%
\be
\{\kappa_1,\kappa_2\}= 2M\Omega\,,\qquad 
\{\kappa_3,\kappa_4\} = -2M\Omega\,,\qquad \{\kappa_5,\kappa_6\}= M\,.
\label{central}
\ee
%%%%%

  The centre-of-mass Hamiltonian (\ref{comH}) is given by
%%%%%
\bea
\cH &=& \fft1{2M}\Big[ (\pi_x + M\nu y)^2 + (\pi_y - M\nu x)^2 +\pi_z^2\Big] 
+\ft12 M \omega^2(x^2+y^2+z^2)
\,.\label{Hcom2}
\eea
%%%%%
It can be verified that the Poisson brackets of the Hamiltonian with
the moment maps are given by
%%%%%
\bea
\{ {\cal H}, \kappa_1\} &=& -(\Omega+\nu)\, \kappa_2\,,\qquad
\{ {\cal H}, \kappa_2\} =  (\Omega+\nu)\, \kappa_1\,,\nn\\
\{ {\cal H}, \kappa_3\} &=& (\Omega-\nu)\, \kappa_4\,,\qquad
\{ {\cal H}, \kappa_4\} = - (\Omega-\nu)\, \kappa_3\,,\nn\\
\{ {\cal H}, \kappa_5\} &=& -\omega^2\, \kappa_6\,,\qquad 
\{ {\cal H}, \kappa_6\} = \kappa_5\,.\label{Hkappa}
\eea
%%%%%
Thus the Poisson brackets of the moment maps with each other and with the
Hamiltonian ${\cal H}$ are the same as the Lie brackets of the
generators $\rK_{(\alpha)}$ with each other and with $\rH$, except for the
central terms in (\ref{central}).

  Note that the moment maps are conserved, i.e. 
%%%%%
\be
\fft{d\kappa_{(\alpha)} }{dt} \equiv \fft{\del \kappa_{(\alpha)} }{\del t} +
\{ {\kappa_{(\alpha)}},{\cal H} \} = 0\,.
\ee
%%%%%a

   In what follows, we shall focus our attention on the centre-of-mass
Hamiltonian for the $(x,y)$ plane alone.  Thus we write
$\cH=\cH_\perp +\cH_3$, where, from (\ref{Hcom2}), we have 
%%%%%
\bea
\cH_\perp &=& \fft1{2M}\Big[ (\pi_x + M\nu y)^2 + (\pi_y - M\nu x)^2\Big]
+\ft12 M \omega^2(x^2+y^2)\,,\\
\cH_3 &=& \fft1{2M}\, \pi_z^2 + \ft12 M\omega^2 z^2\,.
\eea
%%%%%
We may write $\cH_\perp$ as
%%%%%
\bea
\cH_\perp&=& 
\fft{\Omega+\nu}{4M\Omega} (\kappa_1^2 +\kappa_2^2) +
     \fft{\Omega-\nu}{4M\Omega} (\kappa_3^2 +\kappa_4^2)\,,\nn\\
&=& \fft{\Omega+\nu}{4M\Omega} \Big[(\pi_x + M\Omega y)^2 +
          (\pi_y-M\Omega x)^2\Big] \nn\\
&&+
    \fft{\Omega-\nu}{4M\Omega} \Big[(\pi_x - M\Omega y)^2 +
          (\pi_y +M\Omega x)^2\Big]\,.
\eea
%%%%%
It will be convenient to define the complex combinations
%%%%%
\bea
a&=& (\pi_x + M\Omega y) + i (\pi_y-M\Omega x)\,,\nn\\
a^\dagger &=& (\pi_x + M\Omega y) - i (\pi_y-M\Omega x)\,,\nn\\
b&=& (\pi_x - M\Omega y) - i (\pi_y+M\Omega x)\,,\nn\\
b^\dagger &=& (\pi_x - M\Omega y) + i (\pi_y + M\Omega x)\,,
\label{abdef}
\eea
%%%%%
in terms of which we may write the Hamiltonian as
%%%%%
\be
\cH_\perp = \fft{\Omega+\nu}{4M\Omega} \, a^\dagger a +
           \fft{\Omega-\nu}{4M\Omega} \, b^\dagger b\,.
\ee
%%%%%
Note that the angular momentum $J=x \pi_y-y \pi_x$ is given by
%%%%%
\be
J = \fft1{4M\Omega}\, (b^\dagger b - a^\dagger a)\,.
\ee
%%%%%

\section{Quantisation}

   We may pass from the classical algebra of Poisson brackets to the 
quantum commutation relations by means of the standard replacement
%%%%
\be
\bpi  \longrightarrow \hat\bpi = -i\hbar \babla\,.
\ee
%%%%%
One then verifies that with the replacements
%%%%%
\be
\{ A,B\} \longrightarrow \widehat{\{ A,B\} } = \fft{i}{\hbar}\, 
   [ \hat A,\hat B ]\,,
\ee
%%%%
the Poisson bracket algebra given in (\ref{central}) and (\ref{Hkappa}) 
yields the same commutator algebra as that of the quantum operators.

   The quantities $a$ and $b$ defined in (\ref{abdef}) now become the
quantum operators
%%%%%
\bea
\hat a&=& (\hat\pi_x + M\Omega y) + i (\hat\pi_y-M\Omega x)\,,\nn\\
\hat a^\dagger &=&(\hat\pi_x + M\Omega y) - i (\hat\pi_y-M\Omega x)\,,\nn\\
\hat b&=& (\hat\pi_x - M\Omega y) - i (\hat\pi_y+M\Omega x)\,,\nn\\
b^\dagger &=& (\hat\pi_x - M\Omega y) + i (\hat\pi_y + M\Omega x)\,,
\eea
%%%%%
obeying the commutation relations of two mutually commuting sets of 
annihilation and creation operators:
%%%%%
\be
[a,a^\dagger]= [b,b^\dagger]= 4\hbar M \Omega\,,
\ee
%%%%%
and the Hamiltonian and angular momentum can be written as 
%%%%%
\bea
\hat\cH_\perp &=& \fft{\Omega+\nu}{4M\Omega} \, \hat a^\dagger \hat a +
           \fft{\Omega-\nu}{4M\Omega} \, \hat b^\dagger \hat b\
    + \hbar \Omega\,,\\
J &=& \fft1{4M\Omega}\, (\hat b^\dagger \hat b - \hat a^\dagger \hat a)\,.
\eea
%%%%%

   There is therefore 
a unique ground state $|0\rangle$ of energy $\hbar \Omega$  
satisfying $a|0\rangle =0$, 
$b|0\rangle=0$, with excited states of the form
%%%%%
\be
|p,q\rangle = (a^\dagger)^p\, (b^\dagger)^q\, |0\rangle\,,
\ee
%%%%%
and energies
%%%%%
\be
E_{p,q} = (p+q+1)\hbar\Omega + (p-q)\hbar\nu\,.
\ee
%%%%%
The state $|p,q\rangle$ has angular momentum
%%%%%
\be
J_{p,q} = (q-p)\hbar\,.
\ee
%%%%%
These results coincide with those of \cite{Fock} and \cite{Darwin}.

   Concretely, if we write $\zeta=x+i y$, then (setting $\hbar=1$ for 
convenience)
%%%%%
\bea
 \hat a&=& -2i(\bar\del + \ft12M\Omega \zeta)\,,\qquad 
\hat a^\dagger = -2i(\del - \ft12M\Omega \bar\zeta)\,,\nn\\
 \hat b&=& -2i(\del + \ft12M\Omega \bar\zeta)\,,\qquad
\hat b^\dagger = -2i(\bar\del - \ft12M\Omega \zeta)\,,
\eea
%%%%%
where $\del\equiv \del/\del\zeta$ and $\bar\del\equiv \del/\del\bar\zeta$.
The wave functions are of the form
%%%%%
\be
\Psi_{p,q} \propto \bar\zeta^p\, \zeta^q \, e^{-\ft12 M\Omega \zeta\bar\zeta}
\,.
\ee
%%%%%%
The angular momentum operator now takes the form
%%%%%
\be
J= \zeta\del -\bar\zeta\bar\del\,.
\ee
%%%%%

\section{The Lightlike Lift}

   By a result of Eisenhart \cite{Eisenhart}, a classical 
Hamiltonian of the form
%%%%
\be
\cH= \fft1{2m}\, g^{IJ}\, (\pi_I - e A_I)(\pi_J-e A_J) + V(x)
\ee
%%%%%
may be obtained by reduction from the Hamiltonian
%%%%%
\be
\widetilde \cH = g^{IJ}\, \Big(\pi_I -\fft{e}{m}\, A_I \pi_s\Big)
       \Big(\pi_J -\fft{e}{m}\, A_J \pi_s\Big)  + 2 \pi_s \pi_t 
 +\fft{2}{m}\, V\, \pi_s^2\,,
\ee
%%%%%
for a massless particle moving in the higher-dimensional Lorentzian metric
%%%%%
\be
d\tilde s^2 = g_{IJ} dx^I dx^J + \fft{2e}{m}\, A_I dx^I dt + 2 dt dv 
- \fft2{m} V\, dt^2\,.\label{himet}
\ee
%%%%%
The procedure is to impose the massless condition $\widetilde \cH=0$ and 
substitute for the constant component of momentum $\pi_s$ associated to
the null Killing vector $\del/\del v$ the value $\pi_v=m$, and to identify the
moment map $\pi_t$ associated to time translations with $-\cH$.  

    At the quantum
level, the Schr\"odinger equation associated with the quantised Hamiltonian
$\hat H$ can be derived from the massless 
Klein-Gordon equation $\tilde\square\Phi=0$
in the higher-dimensional metric (\ref{himet}), by writing
%%%%%
\be
\Phi(x,t,v) = e^{imv}\, \Psi(x,t)\,,
\ee
%%%%
thus giving
%%%%%
\be
i \fft{\del\Psi}{\del t} =
  -\fft1{2m}\, (\nabla_I - i e A_I)(\nabla_J-i e A_J)\Psi + V\Psi
\ee
%%%%%
in the lower dimension.

  In our case $g_{IJ}=\delta_{IJ}$, $x^I=x_{ia}$ and 
%%%%%
\be
V = \ft12 m \omega^2 \sum_a \br_a^2 + \sum_{a,b} V(\br_a-\br_b)\,.
\ee
%%%%%
The metric (\ref{himet}) can be written as $d\tilde s^2= ds^2 + ds^2_{\rm rel}$,
where
%%%%%
\bea
ds^2 &=& N\Big[(dx-\nu y dt)^2 +  (dy+ \nu x dt)^2 + dz^2 -
              [\Omega^2 (x^2+y^2) +\omega^2 z^2)] dt^2\Big] \nn\\
&&+ 2 dt dv\,,\\
&&\nn\\
ds^2_{\rm rel} &=& \fft1{N}\sum_{a<b}\Big[
 (dx_a-dx_b -\nu(y_a-y_b) dt)^2 + (dy_a-dy_b +\nu(x_a-x_b) dt)^2 \nn\\
&&+
  (dz_a-dz_b)^2 -\{\Omega^2 (x_a-x_b)^2 + \Omega^2(y_a-y_b)^2 
     +\omega^2(z_a-z_b)^2 \nn\\
&&\qquad\qquad + \fft{2}{m} V(\br_a-\br_b)\}dt^2\Big]\,.\label{relmet}
\eea
%%%%%
Note that although we have written the metric as the sum of two terms, it
is not a product metric.  It could be made into a product metric at the
expense of introducing another timelike coordinate $\tau$, so that $t$
in (\ref{relmet}) is replaced by $\tau$.  If $V(\br_a-\br_b)$ is positive,
then $ds^2$ and $ds^2_{\rm rel}$ would then both be Lorentzian.

   The centre of mass metric $ds^2$, and indeed the entire metric 
$d\tilde s^2$, is invariant under the action of the Killing vectors 
$\widetilde \rK_{(\alpha)}$, $\widetilde \rH$ and $\rJ$ which are
the lifts of the vector fields $\rK_{(\alpha)}$, $\rH$ and $\rJ$ that we
introduced in section 2, where
%%%%%
\bea
\widetilde \rK_1 &=& \rK_1 + N\Omega\, [x \sin(\Omega+\nu)t + y
  \cos(\Omega+\nu) t\, ]\, \fft{\del}{\del v}\,,\nn\\
\widetilde \rK_2 &=& \rK_2 +  N\Omega\,[y \sin(\Omega+\nu)t -x 
  \cos(\Omega+\nu) t\,]\, \fft{\del}{\del v} \,,\nn\\
\widetilde \rK_3 &=& \rK_3  + N\Omega\,[x \sin(\Omega-\nu)t - y
  \cos(\Omega -\nu) t\,]\, \fft{\del}{\del v}  \,,\nn\\
\widetilde \rK_4 &=& \rK_4 + N\Omega\,[y \sin(\Omega-\nu)t + x
  \cos(\Omega -\nu) t\,]\, \fft{\del}{\del v}  \,,\nn\\
\widetilde \rK_5 &=& \rK_5 + N\omega z \sin\omega t\, \fft{\del}{\del v}  
  \,,\nn\\
\widetilde \rK_6 &=& \rK_6 - N z \cos\omega t\, \fft{\del}{\del v}  \,,\nn\\
\widetilde\rH &=& \rH\,,\nn\\
\widetilde\rJ &=& \rJ\,.
\eea
%%%%%
The Lie brackets of $\widetilde \rH$ and $\widetilde\rJ$ with 
$\widetilde\rK_{(\alpha)}$ are the same as in (\ref{HKalg}) and (\ref{JKcom}),
but the $\widetilde \rK_{(\alpha)}$ no longer commute with each other.
Their Lie brackets 
%%%%%
\be
[ \widetilde \rK_1,\widetilde \rK_2 ]= -2N \Omega \fft{\del}{\del v}\,,\qquad
[ \widetilde \rK_3,\widetilde \rK_4 ]= 2N \Omega \fft{\del}{\del v}\,,\qquad
[ \widetilde \rK_5,\widetilde \rK_6 ]= -N \fft{\del}{\del v}
\ee
%%%%%
coincide with the negatives of the Poisson brackets of the
moment maps $\kappa_{(\alpha)}$, as given in (\ref{central}).  In fact, one may
pass from the moment maps $\kappa_{(\alpha)}$ to the Killing vectors
$\widetilde\rK_{(\alpha)}$  by means of the
replacements
%%%%%
\be
 \{\pi_x,\pi_y,\pi_z,m\}\longleftrightarrow \Big\{ \fft{\del}{\del x},
   \fft{\del}{\del y},\fft{\del}{\del z}, \fft{\del}{\del v} \Big\}\,.
\ee 
%%%%%

In accordance with Larmor's theorem, the apparent 
dependence of the  metrics (\ref{relmet}) on the Larmor frequency 
$\nu$ may be eliminated  and $\nu$ set to zero, by the 
coordinate transformation (\ref{coord})
with  $\varpi=-\nu$.  The centre of mass metric restricted
to $z=0$ then takes the simple form
%%%%%
\ben
 ds^2 = N\Big[dx^2 +  dy^2 + 
              (\Omega^2 (x^2+y^2) +\omega^2 z^2)dt^2\Big] + 2 dt dv\,.
\een
%%%%%
We shall discuss this further in the next section.

\section{Brdi\v{c}ka-Eardley-Nappi-Witten waves}

In this section we shall relate the Eisenhart lift we have found 
above to  the general class of 
 pp-wave solutions to the Einstein-Maxwell equations that is defined by two 
arbitrary holomorphic functions, $f(u,\zeta)$ and $\phi(u,\zeta)$, 
with complex conjugates $\bar{f}(u,\bar{\zeta})$ and 
$\bar{\phi}(u,\bar{\zeta})$. The configuration takes the form
%%%%%
\bequ
ds^2=du\Big(dv+H(u,\zeta,\bar{\zeta}) du\Big)+
  d\zeta d\bar{\zeta}, \ \ \ \ \  F=
du\wedge(\partial_{\zeta}\phi d\zeta+\partial_{\bar{\zeta}}\bar{\phi} 
d\bar{\zeta}),
\label{ppwave}
\eequ
%%%%
where
\bequ
H(u,\zeta,\bar{\zeta})=f(u,\zeta)+\bar{f}(u,\bar{\zeta})-2\phi(u,\zeta)
\bar{\phi}(u,\bar{\zeta}),
\eequ
and
%%%%
\bequ
F\equiv F_{\mu \nu}dx^{\mu}\wedge dx^{\nu}\,.
\eequ
%%%%%
We have used light-cone coordinates $(u,v)$ and complex coordinates 
$(\zeta,\bar{\zeta})$. The covariantly-constant null vector field is 
$l^{\mu}\partial_{\mu}=\partial / \partial v$. 

A very special property of this solution is that the Maxwell and 
Riemann tensors can be written as
%%%%%
\bequ
F^{\mu \nu}=l^{[\mu}s^{\nu]}, \ \ \ \ R^{\mu \nu \alpha \beta}=l^{[\mu}k^{\nu][\beta}l^{\alpha]}.
\eequ
%%%%%
The vector $s^{\mu}$ and symmetric tensor $k^{\mu \nu}$ are orthogonal 
to $l^{\mu}$. 
Therefore the solution will solve virtually any effective action with 
higher-order curvature terms. This includes closed-string corrections of 
the form $(R_{\mu \nu \alpha \beta})^n$, open-string corrections  of 
the Born-Infeld type and mixed ``corrections'' introducing non-minimal 
coupling. The solutions might however be renormalized by higher-derivative 
corrections. For the gravitational or closed-string corrections this has 
been discussed in  \cite{Horowitz:95}.

A weaker constraint that implies vanishing of some closed-string corrections 
is the one of conformal flatness. The Weyl tensor vanishes for 
(\ref{ppwave}) as long as $\partial_{\zeta}^2 H=\partial_{\bar{\zeta}}^2H=0$ 
(compare with the vacuum equation $\partial_{\zeta}\partial_{\bar{\zeta}}H=0$). 
This is exactly what we will find for the Nappi-Witten case below. It is 
then straightforward to find a set of coordinates where conformal flatness 
becomes explicit.

Some special members of this family of pp-waves are supersymmetric. The Killing spinor equation of ungauged $N=2,D=4$ supergravity,
%%%%%
\bequ
\left(d+\frac{1}{4}\omega_{ab}\Gamma^{ab}-\frac{1}{4}F_{ab}
\Gamma^{ab} \Gamma\right)\epsilon=0\,,
\eequ
%%%%%
reduces to
%%%%
\bequ
\Gamma^{u}\epsilon =0, \ \ \ \ \ 
\partial_{u}\epsilon+2F_{u x}\Gamma^x\epsilon=0\,.
\eequ
%%%%%
We have changed to real coordinates in the transverse space, 
$\zeta=x+iy$. These gamma matrices are flat, i.e. they obey a 
Clifford algebra given by (\ref{ppwave}) with $H=0$. The first condition 
is the usual supersymmetry condition for non-electromagnetic pp-waves. 
The second can only be solved if $F_{ux}$ is independent of the 
transverse space, i.e. if $F_{ux}=f(u)$. 
Then we find the non-trivial Killing spinor
%%%%%
\bequ
\epsilon=e^{-2\int f(u)du \Gamma^x}\epsilon_0, \ \ \ \Gamma^u\epsilon_0=0\,.
\eequ
%%%%%
This wave preserves half of the vacuum supersymmetries. Notice that all 
supersymmetric solutions with $f(\zeta,u)$ linear in $\zeta$ are 
conformally flat.

A slightly more restrictive requirement is 
that the Maxwell field $F_{\mu \nu}$ be covariantly constant
%%%%%
\ben
\nabla_ \rho F _{\mu \nu} \,. 
\een
%%%%%
 It follows that $F_{ux}=C$, or equivalently 
$\partial_{\zeta}\phi(u,\zeta)=C$, and hence 
%%%%%
\bequ
\phi(u,\zeta)=C\zeta, \ \ \ \ \ 
H(u,\zeta,\bar{\zeta})=f(u,\zeta)+\bar{f}(u,\bar{\zeta})-
       2C^2\zeta\bar{\zeta}\,,
\label{BENW}
\eequ 
%%%%%
where $C$ is a constant. This special solution has been considered by 
several authors. Brdi\v{c}ka \cite{Brdicka:51} seems to have been the 
first to obtain it. Eardley \cite{Eardley:74} obtained it from 
requiring $F_{\mu \nu}$ to be covariantly constant.  Nappi and Witten 
constructed this geometry as the target space of a Wess-Zumino-Witten 
conformal field theory with central charge $c=4$ \cite{Nappi:93}, 
although they did not state it is also a solution to Einstein-Maxwell 
theory. The Nappi-Witten form of the metric can be obtained from the 
solution (\ref{ppwave}) with functions (\ref{BENW}) by considering 
$f(u,\zeta)=0$. Performing the coordinate transformations
%%%%%
\bequ
\zeta=(a_1-ia_2)~e^{iu'/2}, \ \ \ u=\frac{u'}{\sqrt{8} C}, \ \ \ 
v=2\sqrt{8} C\left(v'+\frac{bu'}{2}\right)\,,
\eequ
%%%%%
we obtain the metric form used by Nappi and Witten
%%%%%
\bequ
ds^2=2du' dv' +bdu'^2+da^i da^j\delta_{i j}+
  \epsilon_{ij}a_jda_i du', \ \ \ F=\frac{du'}{\sqrt{2}}\wedge
\left(\cos{\frac{u'}{2}}da_1+\sin{\frac{u'}{2}}da_2\right)\,.
\label{npmet}
\eequ
%%%%%
In what follows we will drop the primes for convenience.

Despite appearances, the spacetime (\ref{npmet}) is completely 
homogeneous. In fact, the metric (\ref{npmet}) is a bi-invariant metric 
on the Cangemi-Jackiw group, which we denote by $\mathcal{CJ}$. This 
group is the universal covering group of a central extension of 
${\Bbb E}(2)$, the three-dimensional isometry group of Euclidean two-space, 
by the additive group of real numbers ${\Bbb R}$:
%%%%%
\bequ
\mathcal{CJ}=\widetilde{G_4}, \ \ \ G_4/{\Bbb R}={\Bbb E}(2)\,.
\eequ  
%%%%%
$G_4$ is of course the central extension of ${\Bbb E}(2)$, and the 
tilde stands for universal covering. We denote the generators of 
$\mathcal{CJ}$, $T_A=\{P_i,J,T\}$, for $i=1,2$, where the first three generate, 
respectively, translations and rotation in Euclidean two-space, and 
the last is the central element. An arbitrary element of the 
$\mathcal{CJ}$ group is represented as
%%%%%
\bequ
g=\exp{(a_iP_i)}\exp{(uJ+vT)}\,.
\eequ
%%%%%
The Lie algebra of the group has non-vanishing commutators
\bequ
\left[P_i,P_j\right]=\epsilon_{ij}T, \ \ \ \ \ \ \ \  
\left[J,P_i\right]=\epsilon_{ij}P_j\,,
\eequ
%%%%%
where $\epsilon_{ij}$ is the Levi-Civita tensor density for Euclidean 
two-space. Since $J$ generates rotations we have $0\le u \le 2\pi$, 
for $G_4$. The other coordinates are unconstrained. Topologically 
therefore $G_4\equiv S^1\times {\Bbb R}^3$. But for 
$\mathcal{CJ}=\widetilde{G_4}$ we have $-\infty \le u \le \infty$, 
and therefore $\mathcal{CJ}$ is topologically ${\Bbb R}^4$, just as 
the pp-wave solution.

At this point we would like to remark that the normal subgroup generated 
by $P_i,T$ is clearly the nilpotent Bianchi II Lie algebra, and  so it is
isomorphic to the Heisenberg algebra. This will be of interest in the dynamical analysis performed below.

We now give the isometry group. It was noted in \cite{Nappi:93} that it has 
dimension seven. Explicitly we compute the Killing vector fields to be:
%%%%%
\bequ
\barr{c}
\L_J\equiv \partial_u, \ \ \ R_T\equiv L_T \equiv \partial_v, 
\\\\  R_i\equiv \partial_i+\frac{1}{2}\epsilon_{ij}a_j\partial_v, \ \ \ R_J\equiv \partial_u-\epsilon_{ij}a_j\partial_i, \\\\
L_i \equiv \cos{u}\partial_i+
\sin{u}\epsilon_{ij}\partial_j+\frac{1}{2}
\left(\sin{u}a_i-\cos{u}\epsilon_{ij} a_j\right)\partial_v\,.
\earr
\eequ
%%%%%
These split into two copies of the $G_4$ Lie algebra. The non-trivial 
commutators are:
%%%%%
\bequ
[L_J,L_i]=\epsilon_{ij}L_j, \ \ [L_i,L_j]=\epsilon_{ij}L_T; \ \ \ \ 
[R_J,R_i]=-\epsilon_{ij}L_j, \ \  [R_i,R_j]=-\epsilon_{ij}R_T\,.
\eequ
%%%%%
The isometry group is therefore $(G_4^L\times G_4^R)/N_1$, 
where $N_1$ is the common centre generated by $T$. The right-invariant 
vector fields, $R_i$, generate the left action of the group on the 
spacetime. Considering an element $g'=\exp{(a_i'P_i)}
\exp{(u'J+v'T)}\in G_4^L$ acting on $g=\exp{(a_iP_i)}\exp{(uJ+vT)}$, 
one obtains the transformation
%%%%%
\bequ
(a_i,u,v)\rightarrow (a_i\cos{u'}-\epsilon_{ij}a_j\sin{u'}+a_i', 
 u+u', v+v'+\frac{1}{2}a_k'\epsilon_{ki}a_i\cos{u'}+
\frac{1}{2}a_i'a_i\sin{u'})\,.
\label{coortran}
\eequ
%%%%%
The right action generated by $g'\in G_4^R$ is obtained by 
interchanging primed and unprimed coordinates on the RHS 
of (\ref{coortran}).

A set of right (left) invariant forms, i.e. dual to the right (left) 
invariant vector fields, is $\{\rho^{a}\} (\{\lambda^{a}\})$, given by:
%%%%%
\bequ
\barr{c}
\displaystyle{\rho^{J}=du, \ \ \  \rho_i=da_i+\epsilon_{ij}a_jdu, \ \ \ 
\rho^T=dv-\frac{1}{2}\epsilon_{ij}a_jda_i-\frac{1}{2}a_ia_idu},
\\\\
\displaystyle{\lambda^J=du, \ \ \ \lambda^T=dv+\frac{1}{2}
\epsilon_{ij}a_jda_i, \ \ \ \lambda_i=\cos{u}da_i+\sin{u}\epsilon_{ij}da_j}\,.
\earr
\eequ
%%%%%
The metric can then be written in the explicitly right-invariant or 
left-invariant form
%%%%%
\bequ
ds^2=\rho^T\rho^J+b\rho^J\rho^J+\rho_i\rho_i=2\lambda^J\lambda^T+
b\lambda^J\lambda^J+\lambda_i\lambda_i\,,
\eequ
%%%%%
showing it is the bi-invariant metric on the group.

\section{Conclusion}

In this paper we have provided a fairly full account
of the symmetries, both classical and quantum, of a non-relativistic 
system of 
charged particles each with the same charge to mass ratio,
moving in a magnetic field and harmonic
trapping potential  and subject to mutual 
interactions
depending only on their separations. The assumption of 
equal charge to mass ratios gives rise to a sort of relativity
principle
in which the   Galilei and Bargmann groups are deformed
to  the Newton-Hooke group. We have described the action of this group
on the non-relativistic spacetime which is the analogue of
Newton-Cartan spacetime.  In this way we have given a group theoretic
interpretation 
to a well known theorem of Kohn. Interestingly, Larmor's theorem
becomes
exact in this situation and we used it to show that what appear to be a
one-parameter family of deformations are in fact isomorphic.     

We have also given a ``relativistic'' description in terms of the
null geodesics in a spacetime admitting a null Killing vector field.
In the planar case we have shown that this spacetime is a pp-wave
and may be thought of as a bi-invariant metric on the Cangemi-Jackiw
group.
    
We have not dwelt in this paper on the possible applications of our
results, but  we would like to remark that in addition
to the quantum  Hall effect and  cyclotron resonances \cite{Kohn}, 
and quantum dots \cite{Gritsev,Geyler}, one might expect 
them to be relevant for Bose-Einstein condensates
and the Gross-Pitaevskii equation \cite{cooper,fetter}, 
and possibly also for the collective model
of nuclear physics (we owe this latter suggestion to David  Khmelnitskii). 
 In the light of current interest in laboratory analogue models which
are able to capture some aspects of general relativity, cosmology
and quantum gravity, this might be worth pursuing, as would possible
implications for the AdS/CFT correspondence.   At the more formal level 
 it would be interesting to explore supersymmetric \cite{Galajinsky1}, 
conformal \cite{Galajinsky2} and non-commutative \cite{gomisnc1,gomisnc2}. 
 extensions of our results.

\section{Acknowledgments}

G.W.G. would like to than David Khmelnitskii
for telling him about Kohn's theorem.  We thank Nigel Cooper, 
Joquim Gomis, 
Peter Horvathy and Mikhail Plyushchay for their comments on a preliminary
version of this paper. The research of C.N.P. is
supported in part by DOE grant DE-FG03-95ER40917.

\end{document}